\documentstyle{article}

\begin{document}

\renewcommand{\theequation}{\thesection.\arabic{equation}}

\textwidth = 15truecm
\textheight = 22truecm

\title{Nonelectromagnetic duality \\
in twisted $N=4$ model}

\author{Pei Wang\thanks{e-mail: wp@nwu.edu.cn} \hspace{1cm} Liu Zhao\\
Institute of Modern Physics, Northwest University, Xian 710069, China}

\date{}
\maketitle

\begin{abstract}

In this paper we discuss the possible existing correlation
functions in the $N=4$ topological model. Due to the
distinguished feature that no anomaly exists in $N=4$
supersymmetric theories, the positive-negative ghost number
balance has to be taken into account while considering
the correlation functions. On restriction to K\"{a}hler
manifolds we may find a perturbative mass term which breaks the
$N=4$ supersymmetry down to $N=1$. In all of these,
a nonelectromagnetic duality plays an important role.
Moreover, to get a computable generating functional
the existence of a proper vanishing theorem is required.

\vspace{2cm}

\end{abstract}



\newpage

\section{Introduction}
\setcounter{equation}{0}

Topological quantum field theories \cite{W1} have been pushed forward
vigorously during these years because of the celebrated
Seiberg-Witten theory \cite{W2,W3}. This beautiful theory achieves the 
strong-weak
duality in the $N=2$ supersymmetric model on the one hand and
provides a powerful tool for testing the differential topological
structure of a manifold on the other \cite{S}.

Among topological field theories the correlation function which
represents the Donaldson invariant is one of the essentials \cite{W4}.
However, at least for the gauge group $SU(2)$, known topological
observables are only those represented by correlation functions
of the fields $\mbox{Tr}(\phi^2)$ (in which $\phi$ is a field of ghost
number $+2$) and its descendent $k$-forms ($0<k\leq 4$),
even when matter fields are presented \cite{La}. For $N=4$ model \cite{7},
in which the anomaly-free feature is a well known fact \cite{8},
one may naively think that no nonvanishing correlation functions
(in vacuum) exist because any matrix element of an operator
of positive ghost number vanishes. However, the $N=4$ model is a
larger model including more fields in its multiplets.
Can we expect some suitable operators of negative ghost number
to balance the positive ghost numbers and hence constitute
some nonvanishing correlation functions as topological observables?
The answer is affirmative. In fact we have a nonelectromagnetic
duality in the underlying model under which the fields of
positive and negative ghost numbers are dual to each other.
This makes it possible to construct nonvanishing correlation
functions between pairs of dual fields with opposite ghost numbers.

By using the Mathai-Quillen formalism \cite{MQ} we have given the
action of $N=4$ Yang-Mills model and related BRST and
anti-BRST transformations \cite{10}. In this paper we would like to discuss the
duality symmetry in more detail and use this symmetry to analyze the
topological polynomial invariants in $N=4$ model.

This paper is organized as follows. In section 2 we make a
brief review about the twisted supersymmetric $N=4$ Yang-Mills
theory with emphasis on the description of  nonelectromagnetic
duality symmetry. Next, we show in section 3 that topological
invariants can be constructed from correlation functions of
a pair of dual fields with opposite ghost numbers. Section 4 is
devoted to the understanding of the correlation functions
on K\"{a}hler manifold and the reduction to $N=1$ theory.
We will perform a mass term perturbation in section 5
and give the total result in section 6. In the end of the paper
we shall supply a short discussion.

\section{The action and nonelectromagnetic duality}
\setcounter{equation}{0}

In ref.\cite{10} the action of $N=4$ topological model has
been derived using the Mathai-Quillen technique with
suitable ``nonminimal gauge fermion'' term. The model we would
like to discuss has the twisted $SU(2)_L \times SU(2)_R' \times
SU(2)_F$ symmetry \cite{7}, where $SU(2)_R'$ is the diagonal
contribution of $SU(2)_R \times SU(2)_I$ and $SU(2)_I
\times SU(2)_F=SO(4) \subset SU(4)$, the global symmetry
group of the $N=4$ supersymmetric Yang-Mills model
(the subscripts $I,F$ are used only for distinguishing
the two $SU(2)$ subgroups). So we have the following
spectrum of particles:

\begin{itemize}
\item Bosons

\begin{displaymath}
(1/2,1/2,0)~A_i~(0),~(0,1,0)~B_{ij}~(0),~(0,0,1)~\phi~(2),~\tilde{\phi}~(-2),~
C(0);
\end{displaymath}

\item Fermions

\begin{eqnarray*}
& &(1/2,1/2,1/2)~\psi_i~(1),~\tilde{\chi}_i
~(-1),~(0,1,1/2)~\chi_{ij}~(-1),~\tilde{\psi}_{ij}~(1),\\
& &(0,0,1/2)~\eta~(-1),~\xi~(1),
\end{eqnarray*}
\end{itemize}

\noindent where the ghost numbers are marked in
the brackets following
the field operators, and for clarity we supply that
$B_{ij},~\chi_{ij}$ and $\tilde{\psi}_{ij}$ are chosen
as anti-selfdual fields. With two auxiliary fields
(anti-selfdual) $H_{ij}~(0),~\tilde{H}_i~(0)$ supplemented
we can write down the BRST and anti-BRST transformations \cite{10},

\begin{eqnarray}
\begin{array}{ll}
$$\delta A^i =\epsilon^A \psi^i_A,$$  &
\psi^i_1= \psi^i,~\psi^i_2=\tilde{\chi}^i,$$ \cr
$$\delta B^{ij} =\epsilon^A \chi^{ij}_A,$$  &
\chi^{ij}_1= \tilde{\psi}^{ij},~\chi^{ij}_2=\chi^{ij},$$ \cr
$$\delta \Phi_{AB} =\frac{1}{2}( \epsilon_A \eta_B
+\epsilon_B \eta_A),$$  &
\Phi_{11}=\Phi^{22}= \phi,~
\Phi_{22}=\Phi^{11}=\tilde{\phi},$$ \cr
&
$$\Phi_{12}=\Phi_{21}=-\Phi^{12}=-\Phi^{21}=C,$$ \cr
&
$$\eta_1=\xi,~\eta_2=\eta,$$ \cr
$$\delta \psi^i_A=\epsilon^B(\epsilon_{AB}
\tilde{H}^i + D^i \Phi_{AB})$$,& \cr
$$\delta \chi^{ij}_A=\epsilon^B(\epsilon_{AB}H^{ij}
+ [ B^{ij},~\Phi_{AB} ] ),$$& \cr
$$\delta \eta_A = \epsilon_B [ \Phi_{AC},~\Phi^{CB} ],$$ & \cr
$$\delta \tilde{H}^i = - \epsilon^C(\epsilon^{AB}
[ \psi^i_A,~\Phi_{BC} ] + \frac{1}{2} D^i \eta_C ),$$ & \cr
$$\delta H^{ij} = - \epsilon^C(\epsilon^{AB} [ \chi^{ij}_A, \Phi_{BC} ]
+\frac{1}{2} [ B^{ij}, \eta_C ]). $$ &
\end{array}
\label{2.1}
\end{eqnarray}

The topological charges are defined as ($f$ represents any one
of the above fields)

\begin{equation}
\delta f = \epsilon^A[Q_A, f], \label{2.2}
\end{equation}

\noindent and the action we found can be represented as

\begin{equation}
S=\{Q_1,~[Q_2,~W ]\} = -\{Q^2,~[Q_2,~W]\}, \label{2.3}
\end{equation}

\noindent where

\begin{eqnarray}
& &W=\frac{1}{2e^2}Tr\{ B^{ij}( 2i F^\dagger_{ij} - H_{ij})
+\frac{2i\alpha}{3} B_{ij} [ B_k^i, B^{jk} ] \nonumber\\
& &~~~~~~+ \psi^i \tilde{\chi}_i - 2 \beta C
[ \tilde{\phi},~\phi ] \}. \label{2.4}
\end{eqnarray}

\noindent Later we will take $\alpha=1$ by means of untwisted
$N=4$ supersymmetric transformation. Now we find immediately that
(\ref{2.1}) and (\ref{2.4}) are invariant under the following duality
transform,

\begin{eqnarray}
& & \epsilon^1 \rightleftharpoons \epsilon^2,\nonumber\\
& & i \rightleftharpoons -i,\nonumber\\
& & A^i \rightleftharpoons A^i,\nonumber\\
& & B^{ij} \rightleftharpoons B^{ij},\nonumber\\
& & C \rightleftharpoons C,\nonumber\\
& & \phi \rightleftharpoons \tilde{\phi},\nonumber\\
& & \psi^i \rightleftharpoons \tilde{\chi}^i,\nonumber\\
& & \chi^{ij} \rightleftharpoons \tilde{\psi}^{ij},\nonumber\\
& & \eta \rightleftharpoons - \xi,\nonumber\\
& & \tilde{H}^i \rightleftharpoons - \tilde{H}^i,\nonumber\\
& & H^{ij} \rightleftharpoons -H^{ij}. \label{2.5}
\end{eqnarray}

\noindent For convenience we will call it Gh-duality, because
the dual fields have opposite ghost numbers under this duality.

Using (\ref{2.1}), eq.(\ref{2.3}) gives the explicit action density

\begin{eqnarray}
& & S= \frac{1}{2e^2}\mbox{Tr} \{ - (H_{ij}-i F^\dagger_{ij}
- i \alpha [B_{ki}, B^k_j])^2 - (\tilde{H}_i - 2i D^j B_{ij})^2  \nonumber\\
& & ~~~~~~-(S_{ij})^2 -(k_i)^2 - \chi^{ij} [ \phi, \chi_{ij} ]
- \tilde{\chi}^i [ \phi, \tilde{\chi}_i ]  \nonumber\\
& & ~~~~~~ - \chi^{ij} ( 4i D_i \psi_j +4i \alpha [B_{ki},
\tilde{\psi}^k_j ] - 2 [C, \tilde{\psi}_{ij} ] + [B_{ij}, \xi ] )  \nonumber\\
& & ~~~~~~- \tilde{\chi}^i (4iD^j \tilde{\psi}_{ij} - 4i [B_{ij}, \psi^j]
-2[C, \psi_i] + D_i \xi)  \nonumber\\
& & ~~~~~~ D^i \phi D_i \tilde{\phi} - \psi^i [ \tilde{\phi}, \psi_i]
+\psi^i D_i \eta   \nonumber\\
& & ~~~~~~- [B^{ij}, \phi ] [B_{ij}, \tilde{\phi}] - \tilde{\psi}^{ij}
[ \tilde{\phi}, \tilde{\psi}_{ij} ] + \tilde{\psi}^{ij} [B_{ij}, \eta]
\nonumber \\
& & ~~~~~~ + \beta([ \tilde{\phi},\phi ]^2 + 4 [C, \tilde{\phi} ]
[ C, \phi] + \eta[ \phi,~\eta] + \xi [ \tilde{\phi}, \xi ]
+2 C \{\eta, \xi \}) \},  \label{2.6}
\end{eqnarray}

\noindent in which
\footnote{Here $F^\dagger_{ij}$ is the anti-selfdual part of $F_{ij}=
\partial_i A_j- \partial_j A_i + [A_i, A_j]$, and Vafa and Witten have
proved that $(D^jB_{ij})^2$ can further be expressed as a scalar curvature
and anti-selfdual part of the Weyl tensor in curved space.}

\begin{eqnarray}
& & S_{ij}=F^\dagger_{ij} - i \varpi [B_{ij}, C]
+ \alpha [ B_{ki}, B^k_j ],  \nonumber\\
& &k_i=2 D^j B_{ij} -i \varpi D_i C   \label{2.7}
\end{eqnarray}

\noindent are modified sections of vector boundle
${\cal M} \times_{\cal G} {\cal V}$ (${\cal M}$--moduli
 space of instanton, ${\cal V}$--fiber,
${\cal G}$--gauge transformation group \cite{7,10}), where $\varpi$ changes from
$+1$ to $-1$ under the symmetry (\ref{2.5}). This sign is only superficial
because Vafa and Witten have shown that the crossing terms in $(S_{ij})^2$
and $(k_i)^2$ cancels each other. Therefore we can write

\begin{equation}
S=-\{Q^1, [Q_1, W]\}=-\{Q^2, [Q_2, W]\}=- \frac{1}{2}\{Q^A, [ Q_A, Q]\}.
\label{2.8}
\end{equation}

\noindent The zero section equations

\begin{equation}
S_{ij}=0,~~~~~k_i=0  \label{2.9}
\end{equation}

\noindent denote the self dual equation of instantons, if suitable
vanishing theorem holds \cite{7}.

\section{Observables and correlation functions}
\setcounter{equation}{0}

We first calculate the stress energy tensor. By definition,
the stress-energy tensor $T_{ij}$ obey the formula

\begin{equation}
\delta_M L = \delta_M \int_M \sqrt{g} S =
\int_M \sqrt{g} \delta_M  \label{3.1}
g^{ij} T_{ij}.
\end{equation}

\noindent Since the variation $\delta_M$ of the metric is independent of
supersymmetry transformation, it commutes with the charges $Q_A$.
Besides note that all of the anti-selfdual fields ($\chi_{ij}$ etc.)
should subject to a constraint which requires that \cite{W1}

\begin{equation}
\delta_M \chi_{ij}=-\frac{i}{2} \epsilon_{ijk'l'} \delta_M g^{k'k} g^{l'l}
\chi_{kl} - \frac{1}{4} ( \delta_M g^{rs} g_{rs}) \chi_{ij}. \label{3.2}
\end{equation}

\noindent In particular, this leads to

\begin{eqnarray*}
\delta_M(\sqrt{g} \chi_{ij}\psi^{ij}) = 0
\end{eqnarray*}

\noindent for any tensor field $\psi^{ij}$. It is then straightforward to
write down the stress-energy tensor

\begin{equation}
T_{ij}=-\frac{1}{2} \{Q^A, [ Q_A, V_{ij}] \},  \label{3.3}
\end{equation}

\noindent in which

\begin{eqnarray}
& &V_{ij}= \frac{2}{\sqrt{g}} \frac{\delta_M(\sqrt{g} W)}{\delta_M g^{ij}}
\nonumber\\
& &~~~~~~ =\frac{1}{e^2}\mbox{Tr}\{ \psi_i \tilde{\chi_j}
-g_{ij}(\psi^k \tilde{\chi}_k - 2\beta C [ \tilde{\phi}, \phi])\}. \label{3.4}
\end{eqnarray}

Witten pointed out that the observables of a topological theory
are those operators in the cohomology of related topological charge \cite{W1}. In
the present case, Witten's results on the Feynman integrations will
be generalized to $\langle Q_A, {\cal O} \rangle = 0$ for any operator
${\cal O}$. Moreover, due to the peculiar form of our $T_{ij}$,
the variation of a nonvanishing path integral under a change in the
metric will be zero if either one of $\{Q_1, {\cal O}\}=0$ or
$\{Q_2, {\cal O} \}=0$ holds (we are only interested in those operators
which do not depend on $g_{ij}$ explicitly). So, for the operators which
are neither explicitly depend on the metric nor $Q_A$ exact
we may construct topological invariants. Looking over the
transformations (\ref{2.1}), we can find the candidate fields with ghost
numbers $+4$ and $-4$ respectively,

\begin{eqnarray}
& & {\cal O}_1(x) \equiv {\cal O}^{(0)}_1(x) = \frac{1}{8\pi^2} \mbox{Tr}
\phi^2(x), \nonumber\\
& & {\cal O}_2(x) \equiv {\cal O}^{(0)}_2(x) = \frac{1}{8\pi^2} \mbox{Tr}
\tilde{\phi}^2(x).  \label{3.5}
\end{eqnarray}

\noindent Other observables (if any) should be constructed from the
descendent operators defined as

\begin{equation}
d {\cal O}^{(k)}_A = \{ Q_A, {\cal O}^{(k+1)}_A \},~~~A=1,2. \label{3.6}
\end{equation}

\noindent To be explicit, we have

\begin{eqnarray}
&
\begin{array}{ll}
$${\cal O}_1^{(1)}(x) = \frac{1}{4\pi^2}\mbox{Tr} (\phi\psi)(x)$$, &
$${\cal O}_2^{(1)}(x) = \frac{1}{4\pi^2}\mbox{Tr} (\tilde{\phi}
\tilde{\chi})(x)$$, \cr
$${\cal O}_1^{(2)}(x) = \frac{1}{8\pi^2}\mbox{Tr} (\psi \wedge \psi
+ 2 \phi F)(x)$$, &
$${\cal O}_2^{(2)}(x) = \frac{1}{8\pi^2}\mbox{Tr} (\tilde{\chi}
\wedge \tilde{\chi} + 2 \tilde{\phi} F)(x)$$, \cr
$${\cal O}_1^{(3)}(x) = \frac{1}{4\pi^2}\mbox{Tr} (\psi \wedge F)(x)$$, &
$${\cal O}_2^{(3)}(x) = \frac{1}{4\pi^2}\mbox{Tr} (\tilde{\chi}
\wedge F)(x)$$,
\end{array}
& \nonumber\\
&{\cal O}_1^{(4)}(x) = \frac{1}{8\pi^2}\mbox{Tr} (F \wedge F)(x)
={\cal O}_2^{(4)}(x)&  \label{3.7}
\end{eqnarray}

\noindent in which $\psi=\psi_i dx^i,~\tilde{\chi}=\tilde{\chi}_i dx^i$
are one forms, and $F=dA+A^2= \frac{1}{2}F_{ij}dx^i \wedge dx^j$
is a two form. As a matter of fact, the above operators are
the group of observables with positive ghost numbers found in ref.\cite{W1}
and another group of observables with negative ghost numbers, and both groups
of observables are governed by the Gh-duality.

It is also pointed out by Witten that the corresponding BRST invariants are

\begin{equation}
I_A(\Sigma) = \int_\Sigma {\cal O}_A^{(k)}(x),  \label{3.8}
\end{equation}

\noindent in which $\Sigma$ is a $k$-dimensional homology cycle and
$I_A(\Sigma)$ depends only on the homology class of $\Sigma$. When
we study the correlation functions on simple connected four-manifolds $M$,
we have only to consider $k=0,2$ invariants. So we may have the general
correlation functions

\begin{equation}
\langle {\cal O}_1(x_1) ... {\cal O}_1(x_r)
I_1(\Sigma_1) ... I_1(\Sigma_s)
{\cal O}_2(x_{r+1}) ... {\cal O}_2(x_{r+r'})
I_2(\Sigma_{s+1}) ... I_2(\Sigma_{s+s'}) \rangle.   \label{3.9}
\end{equation}

\noindent Since there is no anomaly in $N=4$ supersymmetry, the ghost
numbers of observables entering the correlation functions have to be
balanced, i.e. we have to impose

\begin{equation}
4r+2s = 4r'+2s'. \label{3.10}
\end{equation}

The problem is now how to construct the generating functional for
the above correlation functions.

Let us study a simple case as an illustrating example. Under the condition
discussed in \cite{W4} via cluster decomposition, one gets

\begin{equation}
\langle {\cal O}_1(x_1) ... {\cal O}_1(x_r)
{\cal O}_2(x_{r+1}) ... {\cal O}_2(x_{r+r'}) \rangle
=\langle {\cal O} \rangle_\Omega^r \langle {\cal O} \rangle_{\Omega'}^{r'}
\langle 1 \rangle ,   \label{3.11}
\end{equation}

\noindent in which $\Omega$ and $\Omega'$ represent different vacua with ghost
numbers $+4$ and $-4$ respectively. The constraint (\ref{3.10}) leads to $r=r'$.
Similar consideration is applicable to $I_A(\Sigma)$. As a whole, we may have

\begin{eqnarray}
& & \langle \exp \sum_{A=1,2} \left( \sum_{a=1}^{s+s'}
\alpha_{Aa} I_A(\Sigma_a) +\lambda_A {\cal O}_A \right) \rangle \nonumber\\
& & ~~~~~= \sum_{\rho} C_\rho \exp \left.\left( \sum_{A,B} \frac{\eta^{AB}_\rho}{2}
\sum_{a,b} \alpha_{Aa}\alpha_{Bb} \#(\Sigma_a \cap \Sigma_b) + \sum_A \lambda_A
\langle {\cal O}_A \rangle_\rho \right) \right|_0, \label{3.12}
\end{eqnarray}

\noindent in which $\#(\Sigma_a \cap \Sigma_b)$ is the intersection number of
$\Sigma_a$ and $\Sigma_b$, the vacuum expectation values $\eta$ and
$\langle{\cal O}\rangle$, $\alpha$ and $\lambda$ are universal constants,
and the symbol $|_0$ means the restriction (\ref{3.10}).

Since the generating functional is quite complicated, we would like to
calculate its main part of contributions instead. Following the explanations
in the end of Section 2, it is clear that the moduli space ${\cal M}$ of instantons
is a subspace of the moduli space of eq.(\ref{2.9}). Thus the zero modes corresponding
to the moduli space ${\cal M}$ cannot include all the fermions in $N=4$ model.
From the transformation law $\delta A^i = \epsilon^A \psi^i_A$, it is easy to
realize that there are two tangents to ${\cal M}$ we can choose, each corresponds
to a group of zero modes.

One may think that the measure for the
path integral has equal numbers of zero modes for fermions
with positive and negative ghost numbers. We thus can split the
integration measure into two parts, one is a measure with positive
ghost number and the other is one with negative ghost number.
Because of eq.(\ref{3.10}), the total integration measure should have no
zero modes. Therefore, the correlation function in (\ref{3.9}) could be splited
into the following form under a reasonable good approximation,

\begin{equation}
\langle {\cal O}_1(x_1) ... {\cal O}_1(x_r)
I_1(\Sigma_1) ... I_1(\Sigma_s) \rangle_1
\langle {\cal O}_2(x_{r+1}) ... {\cal O}_2(x_{r+r'})
I_2(\Sigma_{s+1}) ... I_2(\Sigma_{s+s'}) \rangle_2,   \label{3.13}
\end{equation}

\noindent in which the action $S$ reduces to a pair of (twisted) $N=2$
actions with fermions $\psi^i$, $\chi^{ij}$ and $\eta$ in one of them
and $\tilde{\chi}^i$, $\tilde{\psi}^{ij}$ and $\xi$ in the other.
Obviously, in doing so we have assumed the existence of vanishing theorem and
also a vanishing nonminimal term.

Now assume that there is a mass gap (later on we shall see that a
mass gap do generate when we perform a mass term perturbation which
breaks the $N=4$ supersymmetry down to $N=1$).
Then either parts of the correlation function (\ref{3.13})
can be expressed through a generating functional,

\begin{equation}
\langle \mbox{exp}\left( \sum_a \alpha_{Aa} I_A(\Sigma_a)
+ \lambda_A {\cal O}_A \right) \rangle_A
=\sum_\rho \mbox{e}^{a_\rho \chi + b_\rho \sigma} \mbox{exp}
\left(\frac{1}{2} \eta_{A\rho} \sum_{a,b} \alpha_{Aa} \alpha_{Ab}
\#(\Sigma_a \cap \Sigma_b) + \lambda_A
\langle {\cal O}_A \rangle_\rho \right),  \label{3.14}
\end{equation}

\noindent where $\chi$ and $\sigma$ are Euler characteristic and signature
respectively. Notice that the two-point function
$\langle I_1(\Sigma_a) I_2(\Sigma_b) \rangle$ would not appear in our approaximation.

\section{K\"{a}hler manifolds and reduction to $N=1$}
\setcounter{equation}{0}

To find the formulation on K\"{a}hler manifold we have first to write down
the untwisted $N=4$ supersymmetric transformations. But there is no known
off-shell formulation without constrained fields \cite{11}. So we start from the
on-shell form \cite{12}

\begin{eqnarray}
& & \delta A^i = - \frac{1}2{}( \bar{\eta}_{j \dot{\alpha}}
\bar{\sigma}^{i \dot{\alpha} \alpha} \psi^j_\alpha +
\bar{\psi}^j_{\dot{\alpha}} \bar{\sigma}^{i \dot{\alpha} \alpha}
\eta_{j \alpha}), \nonumber\\
& &\delta \Phi^{ij} = \bar{\eta}^i_{\dot{\alpha}} \psi^{\dot{\alpha} j}
- \bar{\eta}^j_{\dot{\alpha}} \psi^{\dot{\alpha} i}
+ \epsilon^{ijkl} \eta^\alpha_k \psi_{\alpha l}, \nonumber\\
& &\delta \psi^i_\alpha = \bar{\eta}^{\dot{\alpha}}_j D_{\alpha \dot{\alpha}}
\Phi^{ij} + \frac{1}{4} \eta^{i \beta} \sigma^{jk}_{\alpha \beta} F_{jk}
- \frac{1}{2} \eta^j_\alpha [ \Phi^{ik}, \Phi^\dagger_{kj} ], \nonumber\\
& &\delta \bar{\psi}^i_{\dot{\alpha}} = \frac{1}{4} \bar{\eta}^{i \beta}
\bar{\sigma}^{jk}_{\dot{\alpha} \dot{\beta}} F_{jk} -
\frac{1}{2} \bar{\eta}^j_{\dot{\alpha}} [ \Phi^{ik}, \Phi^\dagger_{kj}]
+ \eta^\alpha_j D_{\alpha \dot{\alpha}} \Phi^{ij}. \label{4.1}
\end{eqnarray}

\noindent The spinor algebra and related notations are adopted from
Wess-Bagger's book \cite{13}. Now, the twisted or topological transformation laws
can be obtained by setting

\begin{equation}
\bar{\eta}_{j \dot{\alpha}}= \epsilon^A \sigma_{j A \dot{\alpha}},
~~~\eta_{i \alpha} = 0. \label{4.2}
\end{equation}

\noindent In fact, using the relations

\begin{eqnarray}
& & \Phi^{ij} = i B^{ij} + \frac{1}{2} \sigma^{ij}_{AB}
\Phi^{AB} = \frac{i}{2}
\bar{\sigma}^{ij}_{\dot{\alpha}\dot{\beta}} B^{\dot{\alpha}\dot{\beta}}
+ \frac{1}{2} \sigma^{ij}_{AB} \Phi^{AB}, \nonumber\\
& & \sigma_{i \alpha \dot{\alpha}} A^i = A_{\alpha \dot{\alpha}}, \nonumber\\
& & \sigma_{iA\dot{\alpha}}\psi^i_\alpha = \psi_{\alpha \dot{\alpha} A}
=\sigma_{i \alpha \dot{\alpha}} \psi^i_A, \nonumber\\
& & \bar{\sigma}_{i \dot{\beta}}^A \bar{\psi}^i_{\dot{\alpha}}=
\bar{\psi}^A_{\dot{\alpha} \dot{\beta}} = \frac{1}{2i}
\chi^A_{\dot{\alpha} \dot{\beta}} + \frac{1}{4}
\epsilon_{\dot{\alpha}\dot{\beta}} \eta^A
= \frac{1}{2i} \bar{\sigma}^{ij}_{\dot{\alpha} \dot{\beta}} \chi^A_{ij}
+\frac{1}{4} \epsilon_{\dot{\alpha} \dot{\beta}} \eta^A,  \label{4.3}
\end{eqnarray}

\noindent these transformation laws coincide with (\ref{2.1}) if the auxiliary
fields are replaced by the following expressions (see eq.(\ref{2.6})),

\begin{eqnarray}
& & \tilde{H}^i = 2i D_j B^{ij}, \nonumber\\
& & H^{ij} = i (F^{ij} + \alpha [ B^i_k, B^{kj}]) \label{4.4}
\end{eqnarray}

\noindent with $\alpha=1$.

When the metric on $M$ under consideration is K\"{a}hler, the holonomy
is $SU(2)_L \times U(1)_R$ instead of $SU(2)_L \times SU(2)_R$,
the two-dimensional representation of $SU(2)_R$ decomposes under $U(1)_R$
into a sum of two one-dimensional representations. We follow Witten to
use type $(0,1)$ for one forms $dx^m \sigma_{m\alpha\dot{2}}$, type
$(1,0)$ for one forms $dx^m \sigma_{m\alpha\dot{1}}$. Similarly we have
(notice that the suffices $\dot{1}$ and $\dot{2}$ are interchanged
in our notations as compared to that of Witten)

\begin{equation}
\eta_{i\alpha}=0,~~\bar{\eta}_{j\dot{2}}=\rho^A_1 \sigma_{jA\dot{2}}
~~\mbox{and}~~\bar{\eta}_{j\dot{1}} = \rho^A_2 \sigma_{jA\dot{1}}. \label{4.5}
\end{equation}

For some reasons argued in ref.\cite{W4}, we consider here only $\rho^A_1$ (and
omit the suffix 1 later) symmetry with $\bar{\eta}_{j\dot{1}}=0$. The
transformation laws are

\begin{eqnarray}
& & \delta A^{\alpha \dot{1}} = \rho^A \psi^{\alpha \dot{1}}_A, \nonumber\\
& & \delta A^{\alpha \dot{2}} = 0, \nonumber\\
& & \delta \Phi_{AB} = \frac{1}{4} (\rho_A \eta_B + \rho_B \eta_A), \nonumber\\
& & \delta B^{\dot{1}\dot{1}}=0,\nonumber\\
& & \delta B^{\dot{1}\dot{2}}= \frac{1}{2} \rho^A \chi^{\dot{1}\dot{2}}_A,\nonumber\\
& & \delta B^{\dot{2}\dot{2}}= \rho^A \chi^{\dot{2}\dot{2}}_A,\nonumber\\
& & \delta \psi^{\dot{1}}_{\alpha A} = i \rho^A D_{\alpha \dot{1}}
B^{\dot{1}\dot{1}},\nonumber\\
& & \delta \psi^{\dot{2}}_{\alpha A} = - \rho^B D_{\alpha \dot{1}}
\Phi_{AB} - i \rho_A D_{\alpha\dot{1}} B^{\dot{1}\dot{2}},\nonumber\\
& & \delta \chi^{\dot{1}\dot{1}}_A = 2i \delta \bar{\psi}^{\dot{1}\dot{1}}_A
= \frac{i}{2} \rho_A [ B_{\dot{2} \gamma}, B^{\gamma \dot{1}} ] + \rho^B
[B^{\dot{1}\dot{1}}, \Phi_{AB}],\nonumber\\
& & \delta \chi^{\dot{2}\dot{2}}_A = 2i \delta \bar{\psi}^{\dot{2}\dot{2}}_A
=i \rho_A \bar{\sigma}^{\dot{2}\dot{2}}_{jk} F^{jk},\nonumber\\
& & \delta (\chi^{\dot{1}\dot{2}}_A +\frac{i}{2} \eta_A)
=2i \delta \bar{\psi}^{\dot{1}\dot{2}}_A
= i \rho_A (\bar{\sigma}^{\dot{1}\dot{2}}_{jk} F^{jk} + \frac{1}{2}
[B^{\dot{1}\dot{1}}, B^{\dot{2}\dot{2}} ] )\nonumber\\
& & ~~~~~~~~~~~~~~
+ i \rho^B (\frac{1}{2} [ \Phi_{AC}, \Phi^C_B ]
- i [B^{\dot{1}\dot{2}}, \Phi_{AB}] )\nonumber\\
& & \delta (\chi^{\dot{1}\dot{2}}_A -\frac{i}{2} \eta_A)
=2i \delta \bar{\psi}^{\dot{2}\dot{1}}_A = 0.  \label{4.6}
\end{eqnarray}

\noindent Inspecting eq.(\ref{4.6}) we find that these formulas are almost
a double copy of the corresponding $N=2$ transformation laws (cf.\cite{W4} eq.(3.13))
if we put $B^{\dot{\alpha}\dot{\beta}}=C(=\Phi_{12})=0$,

\begin{eqnarray}
& & \delta A^{\alpha\dot{1}} = \rho^A \psi^{\alpha \dot{1}}_A,~~~
\delta A^{\alpha\dot{2}}=0, \nonumber\\
& & \delta \Phi_{AA} = \frac{1}{2} \rho_A \eta_A, \nonumber\\
& & \delta \psi_{\alpha \dot{2} A}= 0, \nonumber\\
& & \delta \psi_{\alpha \dot{1} A}= \rho^A D_{\alpha \dot{1}}
\Phi_{AA}, \nonumber\\
& & \delta \bar{\psi}^A_{\dot{\alpha} \dot{1}}
= \frac{1}{2} \rho^A \bar{\sigma}^{jk}_{\dot{\alpha}\dot{1}} F_{jk} -
\frac{1}{4} \epsilon_{\dot{\alpha}\dot{1}} \rho^A [ \Phi_{AA}, \Phi^{AA}], \nonumber\\
& & \delta\bar{\psi}^A_{\dot{\alpha}\dot{2}} =0, \label{4.7}
\end{eqnarray}

\noindent where the summation over $A$ appears only in the first equality.
The conditions we used are nothing but the vanishing theorem and
vanishing nonminimal term,
under which the partition function has been represented as the Euler
characteristics of instanton moduli spaces \cite{7,10}, and the correlation functions
can be divided approximately into two parts as is mentioned in the last section.
Vafa and Witten made an
exhaustive study of the subject,
following whom the existence
of vanishing theorem is much convincible
for gauge group $SU(2)$ or a product of $SU(2)$'s \cite{7}.
Provided vanishing theorem holds we can imitate ref.\cite{W4} step
by step to get the correlation functions on K\"{a}hler manifolds.
For example, the (anti)BRST invariance leads to $F^{0,2}=0$,
the holomorphic structure of the bundle is (anti-) BRST invariant,
and $Q^A_1 \equiv \hat{Q}^A$ corresponding to $\rho^A_1$ is enough in analyzing the
topological correlation functions and so on.

However, before studying the mass perturbation we would like to
describe the more general pattern on how to reduce an $N=4$ multiplet
to the $N=1$ multiplets. A tentative scheme is the following,

\begin{eqnarray}
& &\begin{array}{lllcl}
$$N=4$$ (\mbox{twisted}) & $$\longrightarrow$$ & $$N=2$$
& $$\longrightarrow$$ & $$N=1$$ \cr
& & \mbox{Gauge multiplet} & $$U(1)$$ & \mbox{Gauge} \cr
$$H^{ij}~(0,1,0)$$ & $$(0)$$ & $$H^{ij}~(0,1,0) & 0 &
$$H^{\dot{1}\dot{2}}~(0,1^0,0) $$ \cr
$$\tilde{H}^{i}~(1/2,1/2,0)$$ & $$(0)$$ & $$A^{i}~(1/2,1/2,0) & 0 &
$$A_{\alpha \dot{\alpha}}~(1/2,1/2,0)$$ \cr
$$A^{i}~(1/2,1/2,0)$$ & $$ (0)$$ & $$\phi,\tilde{\phi}~(0,0,1^+ \oplus 1^-)$$
& $$2 \oplus -2$$ & $$\psi^1_{\alpha \dot{2}} \equiv \lambda^1_\alpha~
(1/2,1/2^-,1/2^+) $$ \cr
$$B^{ij}~(0,1,0)$$ & $$ (0)$$ & $$\chi^{ij}~(0,1,1/2^-)$$
& $$-1$$ &
$$\left.\begin{array}{l}
$$\tilde{\psi}^2_{\dot{2}\dot{2}}$$\cr $$\tilde{\psi}^2_{\dot{1}\dot{2}}$$
\end{array}\right\} \equiv \bar{\lambda}^1_{\dot{\alpha}}
\begin{array}{l}
$$(0,1^-,1/2^-)$$\cr $$(0,1^0,1/2^-) \oplus(0,0,1/2^-)$$
\end{array}
$$\cr
$$\phi,\tilde{\phi},C~(0,0,1)$$ & $$(2,-2,0)$$ & $$\psi^i~(1/2,1/2,1/2^+)$$
& $$1$$ & \mbox{Chiral U} \cr
$$\chi^{ij},\tilde{\psi}^{ij}~(0,1,1/2)$$ & $$(-1,1)$$ &
$$\eta~(0,0,1/2^-)$$ & -1 & $$H'(H^{\dot{1}\dot{1}},
H^{\dot{2}\dot{2}})~(0,1^+\oplus1^-,0)$$\cr
$$\psi^{i},\tilde{\chi}^{i}~(1/2,1/2,1/2)$$ & $$(1,-1)$$ &
\mbox{Hypermultiplet} &  & $$\Phi'(\phi,\tilde{\phi})~(0,0,1^+\oplus1^-)$$\cr
$$\eta,\xi~(0,0,1/2)$$ & $$(-1,1)$$ &
\tilde{H}^i~(1/2,1/2,0)$$ & 0 & $$\psi^1_{\alpha\dot{1}}\equiv
\psi_\alpha~(1/2,1/2^+,1/2^+)$$ \cr
& & $$B^{ij}~(0,1,0)$$ & 0 &
$$\left.\begin{array}{l}
$$\tilde{\psi}^2_{\dot{1}\dot{1}}$$\cr $$\tilde{\psi}^2_{\dot{2}\dot{1}}$$
\end{array}\right\} \equiv \bar{\psi}_{\dot{\alpha}}
\begin{array}{l}
$$(0,1^+,1/2^-)$$\cr $$(0,1^0,1/2^-) \oplus(0,0,1/2^-)$$
\end{array}
$$\cr
& & $$C~(0,0,1^0)$$ & 0 & \mbox{Chiral V} \cr
& & $$\tilde{\psi}^{ij}~(0,1,1/2^+)$$ & 1 &
$$H''(\tilde{H}_{\alpha\dot{1}})~(1/2,1/2^+,0)$$ \cr
& & $$\tilde{\chi}^{i}~(1/2,1/2,1/2^-)$$ & $$-1$$ &
$$\Phi''(B^{\dot{1}\dot{1}}, B^{\dot{2}\dot{2}})~(0,1^+ \oplus 1^-, 0)$$ \cr
& & $$\xi~(0,0,1/2^+)$$ & 1 &
$$\psi^2_{\alpha\dot{1}} \equiv \tilde{\chi}_\alpha ~(1/2,1/2^+,1/2^-)$$ \cr
& & & &
$$\left.\begin{array}{l}
$$\tilde{\psi}^1_{\dot{1}\dot{1}}$$\cr $$\tilde{\psi}^1_{\dot{2}\dot{1}}$$
\end{array}\right\} \equiv \bar{\tilde{\chi}}_{\dot{\alpha}}
\begin{array}{l}
$$(0,1^+,1/2^+)$$\cr $$(0,1^0,1/2^+) \oplus(0,0,1/2^+)$$
\end{array}
$$\cr
& & & & \mbox{Chiral T} \cr
& & & & $$H'''(\tilde{H}_{\alpha\dot{2}})~(1/2,1/2^-,0)$$\cr
& & & & $$\Phi'''(B^{\dot{1}\dot{2}}, C)~(0,,1^0,0) \oplus(0,0,1^0)$$\cr
& & & & $$\psi^2_{\alpha\dot{2}} \equiv \lambda^2_\alpha~(1/2,1/2^-,1/2^-)$$\cr
& & & & $$\left.\begin{array}{l}
$$\tilde{\psi}^1_{\dot{2}\dot{2}}$$\cr $$\tilde{\psi}^1_{\dot{1}\dot{2}}$$
\end{array}\right\} \equiv \bar{\lambda}^2_{\dot{\alpha}}
\begin{array}{l}
$$(0,1^-,1/2^+)$$\cr $$(0,1^0,1/2^+) \oplus(0,0,1/2^+)$$
\end{array}
\end{array}\nonumber \\
& & \label{4.8}
\end{eqnarray}


\noindent in which $\Phi'(\phi,\tilde{\phi})$ etc. show that $\Phi'$
is a complex field made of two real fields $\phi,\tilde{\phi}$, and 
$( ... ) \oplus ( ... )$ indicates one of the linear combinations of two
states or both.

One may give a mass to the hypermultiplet so that the $N=4$ supersymmetry
reduces to $N=1$ through $N=2$. However this way would break the $SU(2)$
symmetry of the three chiral multiplets in $N=1$ declared by Vafa and Witten
\cite{7},
and the Gh-duality would be lost as well. So we shall consider the case
in which the reduction is direct. The symmetry between three chiral multiplets
thus can be preserved and the Gh-duality can also be inherited as
follows. Replace $\Phi'$ and $\Phi''$
by their complex combinations $\phi$ and $\tilde{\phi}$ so that they are
Gh-dual to each other (we use the same symbols $\phi$ and $\tilde{\phi}$
to show that they have the same behavior under the Gh-duality
transform). 
Then there is also a symmetry between the superfields
$U$ and $V$ under the Gh-duality (to avoid unnecessary complexity we
have neglected the variation of auxiliary fields). 

One may think to construct Gh-selfdual and anti-selfdual multiplets out of the 
gauge multiplet and the $T$ multiplet. However they will not bring us with
new infomation. So we prefer to use the gauge multiplet and the $T$ multiplet
and their respective dual multiplets while computing the mass perturbation.

When $B^{\dot{1}\dot{2}}=C=0$, the multiplets in (\ref{4.8}) will degenerate to
two Gh-dual $N=2$ massless multiplets or two groups of $N=1$ Gh-dual
gauge multiplets and chiral multiplets. Especially the Gh-dual complex fields
$\phi$ and $\tilde{\phi}$ are complex conjugate to each other as well.

\section{The mass term}
\setcounter{equation}{0}

Follow the analysis of supersymmetry mentioned above, a mass-like perturbation
term preserving $N=1$ supersymmetry we can add is

\begin{eqnarray}
\Delta L = - \frac{1}{2} \int_M \sqrt{g} d^4x d^2 \theta \mbox{Tr}
(m_1 U^2 + m_2 V^2 +m_3 T^2 ) - h.c. \label{5.1}
\end{eqnarray}

\noindent in which two types of contribution should be considered:

\begin{itemize}
\item The first two terms with $m_1=m_2=m$ while the ``mass'' can be
replaced by a holomorphic $(2,0)$ form on a K\"{a}hler manifold on
which $H^{2,0}(M) \neq 0$. This is perhaps the case of
``trivial embedding'' in ref.\cite{7}.
\item The last term with $m_3$ a holomorphic two from. This looks like
the ``irreducible embedding'' case \cite{7}.
\end{itemize}

\noindent Because we could not construct an observable from the 
superfield $T$ (which could contribute to the partition function) 
with balanced ghost number, we have to consider only the
first case.

By imitating an analogous discussion made by Witten, we can easily
prove that when the desired vanishing theorem holds the mass term of chiral
superfields $U$ and $V$ are equivalent to 

\begin{equation}
\sum_{A=1,2} \sum_{a} \alpha_a I_A(\omega)  \label{5.2}
\end{equation}

\noindent up to $\hat{Q}_A$-exact terms, where $I_A(\omega)=\int_M
{\cal O}^{(2)}_A \wedge \omega$ are our friends, observables for $k=2$,
and $\omega$ is a nonvanishing holomorphic two form related to $m$. In fact,
Witten chose

\begin{equation}
m=\sigma_{mn \dot{2}\dot{2}} \omega_{kl} \epsilon^{mnkl} \label{5.3}
\end{equation}

\noindent so that (let $\Phi_A=U, V$)

\begin{eqnarray}
& & \Delta L = -\frac{1}{2} \sum_A \int_M
\omega_{kl} d x^k \wedge d x^l d^2 z d^2 \theta \mbox{Tr}
\Phi^2_A \nonumber\\
& &~~~~~~ =-\frac{1}{4} \sum_A \int_M \epsilon^{\alpha \beta}
\sigma_{m \alpha \dot{2}} \sigma_{n \beta \dot{2}}
\omega_{kl} dx^m \wedge dx^n \wedge dx^k \wedge dx^l \mbox{Tr}
\Phi^2_A|_{\theta \theta} \nonumber\\
& &~~~~~~ = - \frac{1}{8} \sum_A \int_M \sqrt{g} d^4x \mbox{Tr}
(m\psi_{A\alpha} \psi^{A\alpha} + \bar{m} \bar{\psi}_{A \dot{\alpha}}
\bar{\psi}^{A\dot{\alpha}} + e^2 \sqrt{2} m\bar{m} \mbox{Tr}(\tilde{\phi}\phi),
\label{5.4}
\end{eqnarray}

\noindent which equals to

\begin{eqnarray}
& & -\frac{1}{2} \int \sqrt{g} d^4 x \mbox{Tr} \psi_{A\alpha}
\psi^{A\alpha} \bar{\sigma}_{mn\dot{2}\dot{2} \omega_kl}
\epsilon^{mnkl}+ \{\hat{Q}_A, V^A \} \nonumber\\
& & \sum_A I_A(\omega) + \{ \hat{Q}_A, V^A + ... \} \label{5.5}\\
\end{eqnarray}

\noindent with

\begin{eqnarray}
& & V^1= -\frac{1}{4} \int \sqrt{g} d^4 x \mbox{Tr} \tilde{\phi}
\bar{\psi}^{\dot{\alpha}} \epsilon^{\dot{\beta}\dot{2}}
\bar{\sigma}_{np\dot{\alpha}\dot{\beta}}\bar{\omega}_{kl} \epsilon^{npkl}, \nonumber\\
& & V^2= -\frac{1}{4} \int \sqrt{g} d^4 x \mbox{Tr} \phi
\bar{\tilde{\chi}}^{\dot{\alpha}} \epsilon^{\dot{\beta}\dot{2}}
\bar{\sigma}_{np\dot{\alpha}\dot{\beta}}\bar{\omega}_{kl} \epsilon^{npkl}, \label{5.6}
\end{eqnarray}

\noindent and the last transformation law in (\ref{4.7}) should be changed into

\begin{eqnarray}
& & \delta \bar{\psi}^{\dot{\alpha}}= -\frac{1}{2} e^2 \rho^1 \phi
\epsilon^{\dot{\alpha}\dot{2}} \bar{\sigma}_{np\dot{2}\dot{2}}
\omega_{kl}\epsilon^{npkl}, \nonumber\\
& & \delta \bar{\tilde{\chi}}^{\dot{\alpha}}= -\frac{1}{2} e^2
\rho^2
\tilde{\phi} \epsilon^{\dot{\alpha}\dot{2}} \bar{\sigma}_{np\dot{2}\dot{2}}
\omega_{kl}\epsilon^{npkl}.  \label{5.7}
\end{eqnarray}

\noindent Hence we have

\begin{equation}
L + \Delta L = L+ \sum_A (I_A(\omega) + \{\hat{Q}_A, ...\}). \label{5.8}
\end{equation}

\noindent Considering the requirement of ghost number balance, the perturbed
correlation function becomes

\begin{eqnarray}
& & \langle {\cal O}_1(x_1)...{\cal O}_1(x_r) I_1(\Sigma_1) ... I_1(\Sigma_1)
\mbox{e}^{I_1(\omega)}\rangle_1  \nonumber\\
& & \times \langle {\cal O}_2(x_{r+1})...{\cal O}_2(x_{r+r'})
I_2(\Sigma_{s+1}) ... I_2(\Sigma_{s+s'})
\mbox{e}^{I_2(\omega)}\rangle_2. \label{5.9}
\end{eqnarray}

\noindent Again, with similar discussions as in \cite{W4}, the expression
for the correlation functions goes back to the key formula (\ref{3.14}).
If the vanishing theorem is valid, the $N=4$ supersymmetry can be
viewed as a pair of $N=2$ supersymmetric theories which further
decompose into $N=1$ ones through perturbation. Consequently we now
have gluino condensation both in regular multiplet and its Gh-dual
multiplet,

\begin{equation}
\langle \lambda_{1\alpha}\lambda_{1}^{\alpha} \rangle =(\mu_1)^3,
\hspace{3mm}
\langle \lambda_{2\alpha}\lambda_{2}^{\alpha} \rangle =(\mu_2)^3,  \label{5.10}
\end{equation}

\noindent where $\mu_1, \mu_2$ are mass scale renormalization parameters.

Since the relavent theories reduce to the minimal $N=1$ systems,
the perturbation leads to a dynamically generated mass gap.

Because of the mass term, the theory now has the $Z_4 \times Z_2'$
symmetry as shown in ref.\cite{W4}, where the $Z_2'$ with generator
$\beta$ is the symmetry which transforms $\Phi_A$ to $-\Phi_A$,
which will be used in the next section.

\section{Polynomial invariants}
\setcounter{equation}{0}

To get the formulae on general K\"ahler manifolds, Witten's routine
requires that one first neglect the difference between
physical and topological theories, i.e. consider the hyper-K\"ahler
case, then make a correction involving the twisting and the
canonical divisor of $M$, when the canonical class of $
M$ is nontrivial.

We first derive the expressions on hyper-K\"ahler manifolds. The
global symmetry group $Z_4 \times Z_2'$ now is broken to $Z_2
\times Z_2'$ with double degeneracy of the vacuum. The two vacuum
states are denoted by $| + \rangle$ and $|- \rangle$. $R$-symmetry
($Z_4$ with generator $\alpha$) tells us that for both $A=1,2$,

\begin{equation}
\alpha: \psi_{A\alpha} \rightarrow i \psi_{A\alpha},~~~
\bar{\psi}_{\dot{\alpha}}^{A} \rightarrow -i \bar{\psi}^A_{\dot{\alpha}}
,~~~{\cal O}_A \rightarrow - {\cal O}_A,~~~
\eta_{A} \rightarrow -\eta_A.     \label{6.1}
\end{equation}

\noindent The last transition is due to that only the $I_A^{(1,1)}$
part of $I_A$ contributes \cite{W4}. The zero modes of $\psi_A$ minus
$\bar{\psi}^A$ is

\begin{equation}
\Delta = 4(k-\nu)+ \nu = \frac{1}{2} \mbox{dim} {\cal M}, \label{6.2}
\end{equation}

\noindent in which

\begin{equation}
\nu \equiv \frac{\chi+ \sigma}{4} =\mbox{integer on the K\'ahler manifolds.}
\label{6.3}
\end{equation}

\noindent Due to the opposite ghost number in the two parts of
the correlation function, we know

\begin{equation}
C_-^1 = i^\Delta C_+^1 = i^\nu C_+^1, ~~ C_-^2 = i^{-\Delta} C_+^2=i^{-\nu}
C_+^2.     \label{6.4}
\end{equation}

\noindent In addition, the canonical divisor on hyper-K\"ahler manifolds
vanishes,

\begin{equation}
K \bullet K= 2\chi+ 3 \sigma =0. \label{6.5}
\end{equation}

\noindent So we can choose

\begin{equation}
C_+^A=\mbox{e}^{a_A \nu}. \label{6.6}
\end{equation}

\noindent As a result, the generating functional for the correlation function
can be written as (set $\eta_A \equiv \eta_{A+}, o_A=
\langle{\cal O}_A \rangle_+$)

\begin{eqnarray}
& & \langle \mbox{exp} \sum_A \left( \sum_a \alpha_{A\alpha} I_A
(\Sigma_a) + \lambda_A{\cal O}_A \right) \rangle  \nonumber\\
& &~~~~=\mbox{e}^{(a_1+a_2)\nu}
\left\{\mbox{exp}\left(\frac{1}{2}\eta_1 \sum_{a,b} \alpha_{1a}
\alpha_{1b}\#(\Sigma_a \cap \Sigma_b) + \lambda_1 o_1 \right) \right. \nonumber\\
& & ~~~~+ \left. i^\nu \mbox{exp}\left(- \frac{1}{2}\eta_1 \sum_{a,b}
\alpha_{1a} \alpha_{1b}\#(\Sigma_a \cap \Sigma_b) - \lambda_1 o_1 \right) \right\}
\nonumber\\
& & ~~~~\times \left\{ \mbox{exp}\left(\frac{1}{2}\eta_2 \sum_{a,b}
\alpha_{2a}\alpha_{2b}\#(\Sigma_a \cap \Sigma_b) + \lambda_2 o_2 \right)
\right. \nonumber\\
& & ~~~~+ \left. \left. i^{-\nu} \mbox{exp} \left(-\frac{1}{2}\eta_2\sum_{a,b}
\alpha_{2a} \alpha_{2b}\# (\Sigma_a \cap \Sigma_b) - \lambda_2 o_2 
\right) \right\} \right|_0.
\label{6.7}
\end{eqnarray}

Next we consider the general K\"ahler case. By general case we mean the K\"ahler
manifold on which $H^{2,0}\neq 0$ and nonzero canonical divisor $C$ exists.
Assume also $C=\bigcup_y C_y$ is the union of smooth, disjoint Riemann surfaces
$C_y$ along which $\omega$ has simple zeros. This brings us with the 
contribution of the so-called
cosmic string \cite{W4}.

Because we have the symmetry breaking from $Z_2 \times Z_2'$ down to $Z_2''$, a
diagonal subgroup generated by the operator $(-1)^F$ in which $F$ counts the
fermion number, the vacua $| \pm \rangle$ bifurcate into four states $| \pm, +
\rangle$ and $| \pm,- \rangle$ near the cosmic string \cite{W4}. Meanwhile, there are
contributions of $\sum_y \#(\Sigma \cap C_y) V_{Ay}$ in which $\#(\Sigma \cap
C_y)$ is the algebraic intersection number between $\Sigma$ and $C_y$, $V_{Ay}$
are local operators $V_A$ inserted on $C_y$ which comes from the integration
$I_A(\Sigma)$. Since the operators $V_A$ transform like $I^{(1,1)}_A$ under the
group $Z_4 \times Z_2'$, i.e.$ \alpha V_A = i V_A, \beta V_A = -V_A$, their
expectation values in four vacua are related in the form

\begin{eqnarray*}
\langle V_A \rangle_{++} =i \langle V_A \rangle_{-+} =-\langle V_A \rangle_{+-}
=-i \langle V_A \rangle_{--} = v_A.
\end{eqnarray*}

\noindent Moreover, the partition functions acquire one more factor \cite{7},

\begin{equation}
\mbox{e}^{\frac{b_A}{2} \chi(C_y)} =\mbox{e}^{b_A(1-g_y)} =
\mbox{e}^{-b_A(2\chi+3\sigma)(M)}, \label{6.8}
\end{equation}

\noindent in which $g_y$ is the genus of $C_y$.

In conclusion, to the approximation we have used, the final generating functional
of $N=4$ topological model on K\"ahler manifold reads

\begin{eqnarray}
& & \langle \mbox{exp} \sum_A \left( \sum_a \alpha_{A\alpha} I_A
(\sigma_a) + \lambda_A{\cal O}_A \right) \rangle \nonumber\\
& &~~~~=\mbox{e}^{(a_1+a_2)\nu}
\left\{\mbox{exp}\left(\frac{1}{2}\eta_1 \sum_{a,b} \alpha_{1a}
\alpha_{1b}\#(\Sigma_a \cap \Sigma_b) + \lambda_1 o_1 \right) \right. \nonumber\\
& & ~~~~~~~~ \times
\prod_y \mbox{e}^{b_1(1-g_y)} (\mbox{e}^{\Phi_{1y}} + t_y \mbox{e}^{-\Phi_{1y}})
\nonumber\\
& & ~~~~+ i^\nu \mbox{exp}\left(- \frac{1}{2}\eta_1 \sum_{a,b}
\alpha_{1a} \alpha_{1b}\#(\Sigma_a \cap \Sigma_b) - \lambda_1 o_1 \right)
\nonumber\\
& & ~~~~~~~~ \times \left.
\prod_y \mbox{e}^{b_1(1-g_y)} (\mbox{e}^{-i\Phi_{1y}} + t_y \mbox{e}^{i\Phi_{1y}})
\right\} \nonumber\\
& & ~~~~\times \left\{ \mbox{exp}\left(\frac{1}{2}\eta_2 \sum_{a,b}
\alpha_{2a}\alpha_{2b}\#(\Sigma_a \cap \Sigma_b) + \lambda_2 o_2 \right) \right.
\nonumber\\
& & ~~~~~~~~ \times
\prod_y \mbox{e}^{b_2(1-g_y)} (\mbox{e}^{\Phi_{2y}} + t_y \mbox{e}^{-\Phi_{2y}})
\nonumber\\
& & ~~~~+ i^{-\nu} \mbox{exp} \left(-\frac{1}{2} \eta_2 \sum_{a,b}
\alpha_{2a} \alpha_{2b}\# (\Sigma_a \cap \Sigma_b) - \lambda_2 o_2 \right)
\nonumber\\
& & ~~~~~~~~ \times \left. \left.
\prod_y \mbox{e}^{b_2(1-g_y)} (\mbox{e}^{-i\Phi_{2y}} + t_y \mbox{e}^{i\Phi_{2y}})
\right\} \right|_0, \label{6.9}
\end{eqnarray}

\noindent in which

\begin{equation}
\Phi_{Ay}=\sum_a \alpha_{Aa} \#(\Sigma_a \cup C_y) v_A \label{6.10}
\end{equation}

\noindent and

\begin{equation}
t_y=(-)^{\epsilon_y},      \label{6.11}
\end{equation}

\noindent $\epsilon_y =0,1$ depending on whether the spin structure is even or odd,
and \cite{W4}

\begin{equation}
\sum_y \epsilon_y = \Delta \quad \mbox{mod} \quad 2 = \nu \quad \mbox{mod} \quad 2.
\label{6.12}
\end{equation}

\section{Concluding remarks}

We have developed the correlation functions for $N=4$ topological model using
the approximation of factorization. The topological invariants thus obtained
are closely related to the Donaldson invariants (the product of Donaldson invariants).
However, if we go further to calculate the corrections (for example, assume the
vanishing theorem fails to hold on some manifolds), we might find some topological
invariants entirely different from Donaldson. Can such invariants bring us
new information in differential geometry? What are their mathematical implications?
These questions may be interesting for further studies.

Another interesting question is to test the modular properties of the
correlation functions and get more evidences for the $S$-duality
\cite{7,14} when the $\theta$-angle
is taken into consideration. One may guess that all of the universal 
constants in our formulas
should become modular functions of $\tau=\frac{\theta}{2\pi} + \frac{4\pi i}{g^2}$
just like in the partition function analyzed by Vafa and Witten \cite{7}.
But it might be
very difficult to find the constraints imposed by modular transformations.

\end{document}